# THE IMPACT OF CELL SITE RE-HOMING ON THE PERFORMANCE OF UMTS CORE NETWORKS

Ye Ouyang and M. Hosein Fallah, Ph.D., P.E.
Howe School of Technology Management
Stevens Institute of Technology, Hoboken, NJ, USA 07030
`youyang@stevens.edu`
`hfallah@stevens.edu`

### ABSTRACT

*Mobile operators currently prefer optimizing their radio networks via re-homing or cutting over the cell sites in 2G or 3G networks. The core network, as the parental part of radio network, is inevitably impacted by the re-homing in radio domain. This paper introduces the cell site re-homing in radio network and analyzes its impact on the performance of GSM/UMTS core network. The possible re-homing models are created and analyzed for core networks. The paper concludes that appropriate re-homing in radio domain, using correct algorithms, not only optimizes the radio network but also helps improve the QoS of the core network and saves the carriers' OPEX and CAPEX on their core networks.*

## 1. INTRODUCTION

The past few years have seen mobile operators transition to next-generation mobile networks; specifically from third-generation networks (3G) to long term evolution (LTE). Subscriber numbers and network usage are up; and forecasts point to even greater expansion for many years. Mobile operators are challenged to retain existing subscribers, acquire new ones, and manage costs for serving both. However, with increased traffic, introduction of data, rich multimedia services as well as larger service areas, the mobile operators are facing the issue of radio network congestion and confronting the demands for larger service coverage areas.

Normally the solution is the network expansion, increasing the size and capacity of mobile networks by installing more network infrastructure into the existing networks, which is an expensive and human-resource intensive undertaking. Therefore, from the radio network aspect, the best approach to avoid cell congestion or channel blocking due to the subscriber increase is to enhance the radio coverage and capacity via increasing the radio infrastructure such as base stations (BTS) and Base Station Controllers (BSC) in Global System for Mobile Communications (GSM), radio domain, or Node-B and Radio Network Controller (RNC) in the Universal Mobile Telecommunication System (UMTS) radio domain.

Meanwhile, a new problem that exists in the radio network expansion is that the load distribution is uneven across the RNCs or BSCs after new Node-Bs or BTSs are added into radio networks. In other words, the new cell sites (BTSs and Node B) may cause a few BSCs and RNCs to overload. The traffic should be re-distributed across all the old and new BSCs and RNCs. Mobile operators usually strive to resolve this issue by cell site re-homing; move ("re-home") a few cell sites from BSC or RNC with heavy load to that with low load. Generally re-homing is a re-distribution and re-configuration process for traffic and routing in the radio domain. It is of pivotal importance for the mobile operators to optimize the GSM/UMTS radio domain to the extent possible before investing more to expand the infrastructure.

Figure 1 displays the topology of a GSM/UMTS radio domain in which BTS and Node-B are the first reference point for end subscribers to access into the mobile networks. BSC and RNC, standing above cell sites logically, are responsible for the connection, control and management of base stations in the radio domain. In particular, the radio domain in UMTS, according to 3GPP TS 25.401 and 3GPP TS 23.002, is called UMTS Terrestrial Radio Access Network (UTRAN) which includes one or multiple Radio Network Sub-system (RNS). An RNS contains one RNC and one or more Node B. Similarly, the radio domain in GSM is called GSM RAN which includes one or more Base Station Sub-system (BSS). A BSS contains two types of Network Entities: BSC and BTS. BSC plays a similar role as RNC to control and route calls for the base stations. BTS as the cell site in GSM is to access the mobile stations (MS). The re-homing technique towards the radio domain attempts to achieve the optimization of routings, loading and throughput for both 2G and 3G RAN.

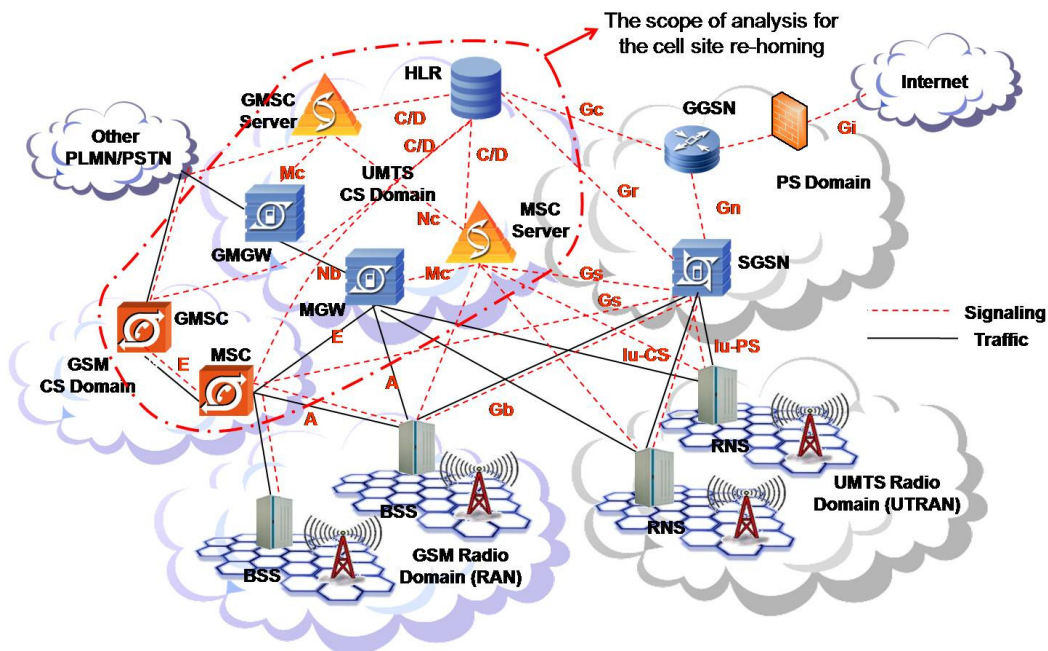

Figure 1. Architecture of the GSM/UMTS network and the impact scope of the cell site re-homing

Oom, Jan. et al (2004) stated that the cell site re-homing procedure for radio access networks (RAN) requires many manual operation steps, so it is a labor intensive and time-consuming task requiring reconfigurations of both the radio and transport networks. Take 2G RAN as a example; a cell site re-homing requires the following operations: 1) verify the hardware configuration in a target BSC by comparing it to the hardware configuration of a source BSC; 2) check if the relevant software versions for BTS are available in the target BSC by comparing them with the registered versions in the source BSC; 3) copy the cell data from the source BSC to the target BSC; 4) copy the site data from the source BSC to the target BSC; 5) copy the neighbor cell data from the source BSC to the target BSC; 6) create new external cells data in the source BSC with state "not operating;" 7) create new external cells data in the target BSC with state "not operating;" 8) halt source cell in the source BSC; 9) block transceiver (TRX) resources in the BTS; 10) set old external cells data in the source and target BSC to state "not operating." With the above preliminary measures completed then; 11) moving the switching connection for the BTS from the source BSC to the target BSC. The re-homing procedure continues with the following operations: 12) update the Cell Global Identifier (CGI) in the Mobile Services Switching Center (MSC); 13) set new external cells data in the source and target BSCs with state "operating;" (14) de-block TRX resources in the BTS; 15) activate target cell in the target BSC; 16) remove the cell data in the source BSC; 17) remove the site data in the source BSC; 18) remove N-cell data in the source BSC; 19) remove old external cells data in the source BSC; and 20) remove old external cells data in the target BSC.

The dot dashed circled area in the Figure 1 shows the impact scope of core networks due to the re-homing in radio access networks (RAN). The core network (CN) is the heart of the current mobile communication networks. CN and RAN are closely coupled via A interface between BSC and MSC in GSM network and via Iu-CS interface between RNC and Media Gateway (MGW) in UMTS network (3GPP TS 25.413and 3GPP TS 25.415). Therefore, the changes in routing, loading and throughput resulting from the re-homing at radio side will definitely impact the performance of core network in GSM or UMTS. As per Figure 1, the cell site re-homing in GSM RAN and UTRAN impacts the area with red dot dashed line circled through A and Iu-CS interfaces. The paper will study the impact of the re-homing in RAN on the performance of CN in GSM/UMTS networks. The appropriate re-homing in radio domain, followed by corresponding regulations, not only optimizes the radio network but also helps improve the QoS of core network and reduces the need for extra OPEX and CAPEX investment in on core network.

The rest of the paper proceeds as follows: Section 2 summarizes the literature in the related area and the challenges in optimizing and planning mobile core networks. Section 3 introduces the proposed re-home models in mobile core networks and how a re-homing in the radio domain impacts the performance of the Core Network (CN). Section 4 which is the core of the paper discusses the numeric analysis for the proposed rehoming models and creates the algorithms to optimize the core networks via rehoming operations. Section 5 provides a case study to illustrate application of the algorithms created in Section 4. Section 6 is the conclusion to the paper.

## 2.  LITERATURE REVIEWS

The current literature is more focused on the practical or theoretical solutions to design and plan GSM, UMTS, and Long Term Evolution (LTE) radio networks but overlooks the algorithms for planning and optimization of core networks. Also earlier studies don't  provide a unified approach to optimize the traffic and throughput for mobile core networks. Furthermore, the current literature does not address  the dimensioning and optimization for mobile core networks. This can be due to: 1) The mobile core network in either logical or physical structure is more complicated than radio access network, 2) The internal traffic and throughput may vary depending on the vendors' network entities (NE).

Oom, Jan. et al (2004) in their patent introduce a resource sharing method through rehoming work in wireless radio networks. The patent interprets the detailed rehoming procedure and achieves the optimization of GSM radio networks via re-balancing and re-distributing the traffic going through the network. Their method is a good starting point for extending the rehoming to UMTS core networks. Shalak, R. et al (2004) make a qualitative study of the performance of UMTS core network, in which equipment of multiple vendors of UMTS CN are compared. Harmatos, J. (2002) proposes a model to plan UMTS core network based on the requirements from radio access network. The model also considers the premise of planning work in cost minimization, which helps mobile operators minimize Capital Expenditure (CAPEX).Their solutions are more based on techno-economic aspect to achieve the maximization of core network performance and minimization of Total Cost of Ownership (TCO) through network planning. Many other articles discussed the planning and dimensioning methods for GSM and UMTS network (Konstantinopoulou, C.N., 2000; Mishra, A , 2003; Britvic, V. et al, 2004; Vrabel, A. et al , 2007; and Szekely, I. et al, 2008). These articles are high level perspectives and interpretation of the on network transition and the evolution of the core network architecture. Ouyang, Y. and Fallah, M. H. (2009) make a further study to extend the mobile network planning and dimensioning into the level of network entities. That is, to study the throughput generated and absorbed in each interfaces of network entities. The algorithms created for the interface level provide a guideline for mobile operators to dimension their UMTS core networks.

We believe the area that most of the current literature has overlooked is the core network optimization. Therefore, the objective of this paper is to analyze the impact of radio network rehoming on the core network domain, to create a unified algorithm to address the traffic/throughput re-distribution and re-configuration in UMTS core networks, and to achieve the throughput/ traffic optimization in core network domain via the rehoming operations between RNC and MGW.

## PROPOSED RE-HOMING MODELS

This section will analyze how a re-homing in the Radio Domain impacts the performance of the Core Network. As per the twenty re-homing steps in section 1, the routing configuration, traffic loading, and data throughput of involved RNC or BSC are changed after a re-homing of related cell sites (Node-B or BTS). The changes in RNC or BSC due to re-homing will result in the new changes in the traffic load and routings in 3G MGW via Iu-CS interface or 2G MSC via A interface. Let's consider a possible scenario: from the radio domain aspect, if a re-homing evenly re-distributes the traffic across all involved RNC and reduces the loading of all involved RNC to a reasonable value (normally below 70% or 80% of the loading threshold of a RNC), it can be recognized as a successful re-homing in the radio domain. However, the re-distributed traffic in a RNC flowing into its paired MGW via Iu-CS interface sometimes may be over the tolerable threshold of the MGW in traffic loading. Hence, this re-homing is not executable from the core network aspect. One simple approach to resolve this issue is to add new MGW to share the load. That will require CAPEX. An alternative is "re-rehoming," to re-distribute the traffic across the involved MGW based on the re-homing for RNC. The intention of the second re-homing is to 1) Ensure the successful implementation of the re-homing in the radio domain; 2) optimize the core networks by re-distributing the traffic across all involved 3G MGW, 3G MSC Server (MSS) or 2G MSC. The effectiveness of re-homing depends on optimization of both the radio and the core networks. The potential re-homing models for the core network are described below.

## 2.1 MODEL 1: FROM SINGLE 3G MGW TO SINGLE 3G MGW

Figure 2 illustrates the re-homing process for mode 1: Two single and neighboring MGWs: MGW A reports to MSS A and MGW B reports to MSS B belong to the same market A. Both MGW A and B are the only media gateway under control by their paired MSS. MSS A and B must belong to the same geographical service area or market. In this model, a RNC or BSC used to connect to MGW A is re-homed to MGW B.

For simplicity of Operation and Maintenance (O&M), the scope of a re-homing is normally in the same service area or market, which means that in model 1 the re-homing is only allowed in Market A but not to extend to other markets. The scope of the re-homing is applied to all the following models in the paper.

Sometimes there are multiple MGWs under control of the same MSS in a service market. From the aspect of re-homing, it is theoretically expected that the traffic distribution across these MGWs will always be even and equal. Therefore, two other principles for re-homing in the core network side should be defined: 1) a re-homing is not allowed to execute from one MGW to another if this source and target MGW belong to the same MSS. 2) If a RNC or BSC is re-homed into a target MSS who controls multiple target MGWs, the re-homed traffic by the RNC/BSC should be evenly distributed into all the target MGWs under the target MSS. The principles are applied to all the re-homing modes.

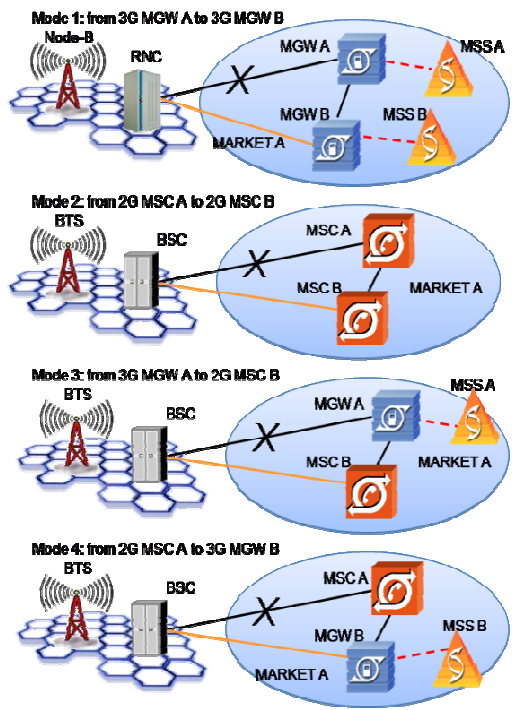

Figure 2. Re-homing model 1, 2 and 3

## 2.2 MODEL 2: FROM SINGLE 2G MSC TO SINGLE 2G MSC

Figure 2 also illustrates the re-homing process for model 2 which is the simplest model: Two single and neighboring MSCs: MSC A and MSC B belong to the same market A. Both MSC A and B belong to the same geographical service area or market. Model 2 is similar to model 1 and complies with the same principles. In this model, a BSC used to connect to MSC A is re-homed to MSC B.

## 2.3 MODEL 3: FROM SINGLE 3G MGW TO SINGLE 2G MSC

Figure 2 also illustrates the re-homing process for model 3: Two single and neighboring MGW A and MSC B: MGW A who reports to MSS A and MSC B belong to the same market A. both MGW A and B belong to the same geographical service area or market. MGW A is the only media gateway under MSS A in market A. In this model, a BSC used to connect to MGW A is re-homed to MSC B.

## 2.4 MODEL 4: FROM SINGLE 2G MSC TO SINGLE 3G MGW

Figure 2 also illustrates the re-homing process for model 4: Two single and neighboring MSC A and MGW B: MSC A and MGW B who reports to MSS B belong to the same market A. Both MGW A and B belong to the same geographical service area or market. MGW B is the only media gateway under MSS B in market A. In this model, a BSC used to connect to MSC A is re-homed to MGW B.

## 2.5 MODEL 5: FROM SINGLE 2G MSC TO MULTIPLE 3G MGW

Figure 3 illustrates the re-homing process for model 5: in this model, MSC A is the source MSC in the re-homing in market A, while MSS B controls MGW B1, B2 to Bn in market A and MGW Bn+1 in market B. This scenario considers that a MSS sometimes is responsible for the management of multiple MGW from different service areas or markets. The single MGW A who reports to MSS A with its multiple neighboring MGW B1, B2 to Bn belong to the same market A. The Inter-Machine Trunks (IMT) are hidden in the figure.

In this model, a BSC used to connect to MGW A is re-homed to all the target MGWs (1,2…n) who are under MSS B and in the market A. As per principle 1 defined in model 1, any MGW under MSS B but belongs to non-market A should not be considered as the target MGW for re-homing. So MGW Bn+1is excluded from the target MGWs for the re-homing. As per principle 2 defined in model 1, the re-homed traffic from the BSC should be evenly distributed into MGW B1, B2… Bn.

## 2.6 MODEL 6: FROM SINGLE 3G MGW TO MULTIPLE 3G MGW

Model 6 is almost the same as model 5. The only difference is the source switch: in model 6, the source switch for re-homing is a single 3G MGW while it's a 2G MSC in model 5. MGW A is the only media gateway under MSS A in market A, while MSS B controls MGW B1, B2 to Bn in market A and MGW Bn+1 in market B. This scenario considers that a MSS sometimes is responsible for the management of multiple MGW from different service area or markets. The single MGW A who reports to MSS A with its multiple neighboring MGW B1, B2 to Bn belong to the same market A. The Inter-Machine Trunks (IMT) are hidden in the figure.

In this model, a RNC or BSC used to connect to MGW A is re-homed to all the target MGWs (1,2…n) who are under MSS B and in the market A. As per principle 1 defined in model 1, any MGW under MSS B but belongs to non-market A should not be considered as the target MGW for re-homing. So MGW Bn+1is excluded from the target MGWs for re-homing. As per principle 2 defined in model 1, the re-homed traffic from the RNC or BSC should be evenly distributed into MGW B1, B2… Bn.

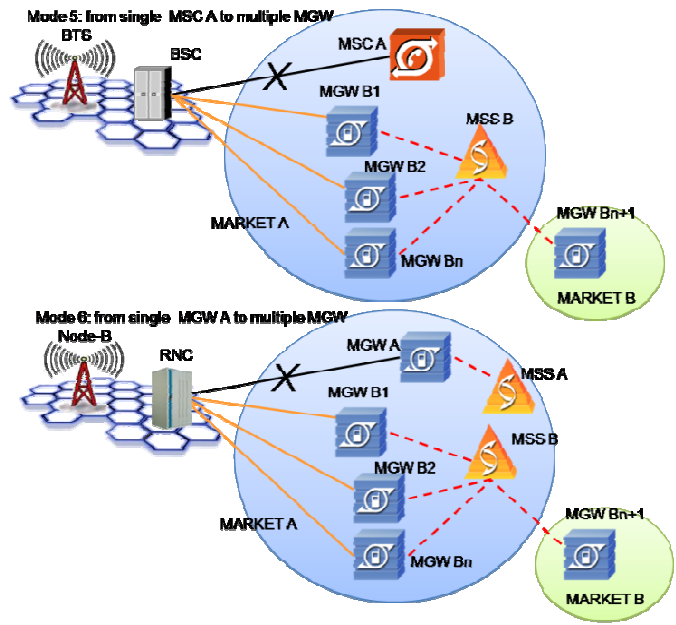

Figure 3. Re-Homing Model 5 And 6

## 2.7 MODEL 7: FROM MULTIPLE 3G MGW TO SINGLE 2G MSC

Figure 4 illustrates the re-homing process for model 7: MSS A controls MGW A1, A2 to An in market A and MGW An+1 in market B. MSC B in market A is the target switch for the re-homing. This scenario considers that a MSS sometimes is responsible for the management of multiple MGW from different service area or markets. The Inter-Machine Trunks (IMT) are hidden in the figure.

In this model, the BSC used to connect to MGW A1, A2…An in market A is re-homed to the target MSC B in market A. As per principle 2 defined in model 1, the total re-homed traffic into MSC B is evenly coming from MGW A1, A2…An.

## 2.8 MODEL 8: FROM MULTIPLE 3G MGW TO SINGLE 3G MGW

Figure 4 also illustrates the re-homing process for model 8: MSS A controls MGW A1, A2 to An in market A and MGW An+1 in market B. MSS B controls MGW B1 in market A and MGW B2 in market C. MGW B1 is the only media gateway in market A under control by MSS B. This scenario considers that a MSS sometimes is responsible for the management of multiple MGW from different service area or markets. The Inter-Machine Trunks (IMT) are hidden in the figure.

In this model, A RNC or BSC used to connect to MGW A1, A2…An in market A is re-homed to MGW B1 in market A. As per principle 1 defined in model 1, any MGW under MSS B but belong to non-market A should not be considered as the target MGW for re-homing. So MGW B2 is excluded from the target MGWs for the re-homing. As per principle 2 defined in model 1, the total re-homed traffic into MGW B1 is evenly coming from MGW A1, A2…An.

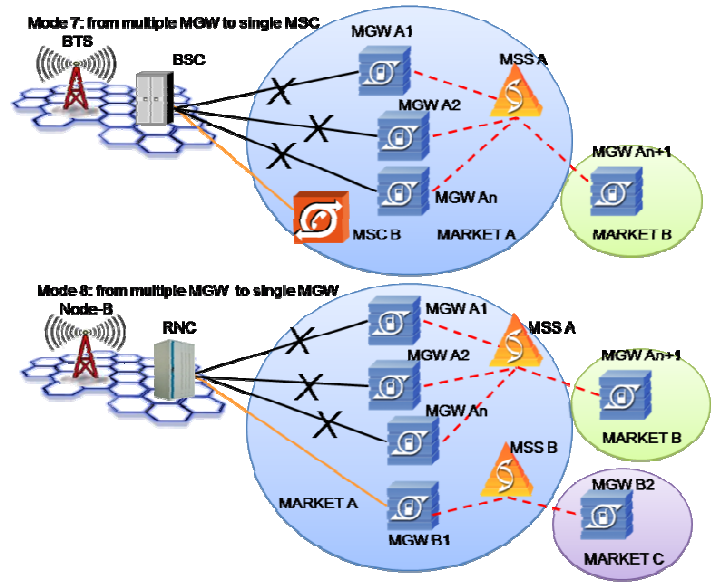

Figure 4. Re-homing model 7 and 8

### 2.9 MODEL 9: FROM MULTIPLE 3G MGW TO MULTIPLE 3G MGW

Figure 5 illustrates the re-homing process for model 9 which is the most complicated scenario: MSS A controls MGW A1, A2 to An in market A and MGW An+1 in market B while MSS B controls MGW B1, B2…Bn in market A and MGW Bn+1 in market C. This scenario considers that a MSS is sometimes responsible for the management of multiple MGW from different service area or markets. The Inter-Machine Trunks (IMT) are hidden in the figure.

In this model, a RNC or BSC used to connect to MGW A1, A2…An in market A is re-homed to MGW B1, B2…Bn in market A. As per principle 1 defined in model 1, any MGW under MSS B but belong to non-market A should not be considered as the target MGW for re-homing. So MGW Bn+1 is excluded from the target MGWs for the re-homing. As per principle 2 defined in model 1, the re-homed traffic from the RNC or BSC should be evenly distributed into MGW B1, B2…Bn.

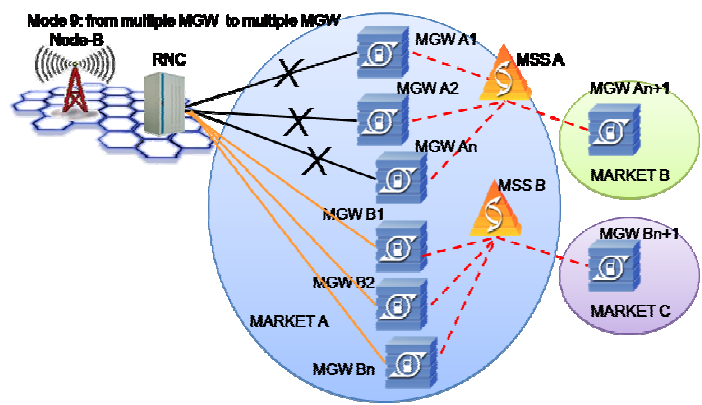

Figure 5. Re-homing model 9

### 3. NUMERICAL ANALYSIS

#### 4.1. PARAMETERS TO CONSIDER IN A RE-HOMING

Since the ultimate goal of the mobile operators is to save the CAPEX and optimize the

56

network via the re-homing in both the radio network side and the core network side, we need to analyze the result of the optimization via re-homing, e.g. how the traffic or routings are re-distributed. Also we may figure out the cost reduction via re-homing such as the cost savings between a normal network expansion and a re-homing. These two results are extremely significant for mobile operators in their strategy making to find a balance between the Quality of Service (QoS) of their networks and the capital investment on the infrastructure. All re-homing scenario in core networks actually can be reduced into one of the nine models summarized in section 2. So the numerical analysis in this section is based on the nine re-homing models.

In the re-homing in core network side, we are focused on the optimization result of the switches (MSC, MGW, and MSS) and the cost reduction on core network expansion (re-homing VS non re-homing). In a mobile switching center, three major parameters are considered: T1 or E1 ports number, Busy Hour Calling Attempt (BHCA) capacity and the utilization of Signaling System #7(SS7). More specifically, the following parameters shall be calculated:

1) Forecasted E1 or T1 ports needed for the switch (Re-homing VS Non re-homing).

The periodic forecasting value (monthly, quarterly, and annually) weighs how many new trunks are needed to configure for the existing switch. Meanwhile it also indicates the rough cost on the expansion for the current switch by multiplying the unit price of T1/E1 lines.

2) Whether a new switch needs to be purchased (Re-homing VS Non re-homing).

Sometimes only new T1/E1 ports added into the current switch can meet the requirement of the increasing traffic at a specific time spot. Sometimes one or more new switches need to be added to share the increasing traffic when the increase in traffic goes beyond the maximum capacity of existing switch.

3) Forecasted total E1/T1 ports needed for the new switch (Re-homing VS Non re-homing)

If one or more new switches are needed, this parameter weighs how many new trunks are needed to be configured for the new switches. Also it indicates the rough cost for the new infrastructure by multiplying the unit price of T1/E1 lines.

4) Forecasted BHCA capacity (Re-homing VS Non re-homing)

The mobile operators ensure the load is below the maximum BHCA capacity of the switch. A redundancy factor normally is considered.

5) Forecasted SS7 signaling utilization (Re-homing VS Non re-homing)

The mobile operators ensure the load is below the maximum SS7 utilization. A redundancy factor is normally considered.

**4.2. FORECASTING THE PERFORMANCE OF SWITCH BEFORE A RE-HOMING**

In any re-homing scenario, the involved switches can be divided into two types: source switches or target switches. In a re-homing, the load of a source switch is always decreased after a re-homing while the load of target switch will be raised. Therefore the performance of the target switches in a re-homing shall be more carefully investigated.

Table 1. Forecasted utilization with or without a re-homing

| Switch | Limits | | | Forecasted Utilization (Before Re-homing) | | | | Forecasted Utilization (After Re-homing) | | | |
|---|---|---|---|---|---|---|---|---|---|---|---|
| | Criteria | Installed Capacity | Maximum Capacity | Month 1 | M 2 | … | M k | Month 1 | M 2 | … | M k |
| Source or Target Switch | BHCA | $A_{IC}$ | $A_{Max}$ | a1 | a2 | … | ak | A1 | A2 | … | Ak |
| | Ports | $B_{IC}$ | $B_{Max}$ | b1 | b2 | … | bk | B1 | B2 | … | Bk |
| | SS7 | $C_{IC}$ | $C_{Max}$ | c1 | c2 | … | ck | C1 | C2 | … | Ck |

Table 1 shows the forecasted utilization of source and target switches before and after a re-homing. The values of forecasted utilization before re-homing can be obtained through the traffic of switch.

$$Y_n = NS \times Erlang_{NS}, \qquad (1)$$

Let NS be the predictor variable thought to be related to a response variable Y where

NS=Number of subscribers in a service area in one specific month (n=1,2…k). It is determined by the development plan of the mobile operators.

$Erlang_{NS}$ denotes the average erlang per subscriber. It is determined by the traffic model defined by the mobile operators.

Yn=Switch traffic in month n(n=1,2…k). In this case, we use Erlang to represent the traffic.

The Yn (traffic of the switch) in Equation 1 will be used to obtain the forecasted utilization including such parameters as the monthly forecasted values of BHCA, T1/E1 ports and SS7 before a re-homing

We have

$$BHCA_{Before} = a_n = CAPS \times 3600 = Erlang \times 3600/T = Y_n \times 3600/T \qquad (2)$$

where $BHCA_{Before}$ denotes the forecasted BHCA before a re-homing.
CAPS denotes the call attempts per second,
Yn denotes the forecasted traffic in month N obtained via formula 5 and 1,
T denotes the average time per call.

The forecasted number of trunks before a re-homing is given by

$$N_{Trunk-Before} = b_n = Erlang/(L_{Channel} \times N_{Channel}) = Y_n/(L_{Channel} \times N_{Channel}) \qquad (3)$$

where $N_{Trunk-Before}$ denotes the forecasted number of trunks before a re-homing,
$L_{Channel}$ represents the loading of traffic channel. Normally it's 0.7 Erlang/Channel,
$N_{Channel}$ denotes the number of channels per trunk. T1 is 24 and E1 is 30.

An alternative to obtain $N_{Trunk-Before}$ is to use Erlang B calculator by inputting the busy hour traffic (BHT) and blocking rate. However, the results obtained from Erlang B calculator are longer appear as a linear relation defined in Equation 1.

### 4.3. FORECASTING THE PERFORMANCE OF SWITCH AFTER A RE-HOMING.

A re-homing theoretically can be executed at any time. However, since the goal of a re-homing is to optimize the network via redistributing the traffic and routings, a preferable time slot for a re-homing is when any value of the forecasted utilization is or will be close to the threshold of installed capacity of the switch.

Assume implementation of a re-homing in month N-1, $(N \geq 2)$ which means that all the forecasted utilization values will be changed and optimized from month N. In the re-homing, assume a RNC is re-homed from a source switch to another target switch. The RNC carries $N_{RNC}$ T1 trunks and $ERL_{RNC}$ erlang traffic.

The forecasted BHCA in month N after a re-homing is provided below:

$$BHCA_{After} = A_n = CAPS \times 3600 = \frac{(Y_n \pm ERL_{RNC}/n) \times 3600}{T}$$
$$= \frac{(Y_n \pm ERL_{RNC}/n) \times 3600}{Y_1 \times 3600/a_1} = a_1 \times \frac{(Y_n \pm ERL_{RNC}/n)}{Y_1} \quad (4)$$

where a1 denotes the forecast BHCA in month 1 before a re-homing,

Y1 denotes the traffic of the switch in month 1,

"-" denotes it's the source switch; "+" denotes it's the target switch,

n denotes the number of source or target switches.

The forecasted number of T1/E1 trunks in month N is given by

$$N_{Trunk-after} = B_n = b_n \pm \left(\frac{N_{RNC}}{n}\right) \quad (5)$$

where $b_n$ denotes the forecasted number of trunks in month N before a re-homing.

### 4.4. ESTIMATION OF THE COST SAVING WITH A RE-HOMING.

Why a re-homing is a cost savings because the original investment on the capacity expansion for the high loading switches is no longer needed or only partially needed. The re-homing unloads the partial or whole traffic from the high loading source switches and moves it to the target switches with lower loading. The traffic re-distribution optimizes the utilization of network capacity. The re-homing enables the whole network to be a real net in which the link connecting each switch can absorb and balance the traffic between high and low loading nodes.

Consider the simplest model 1 in section 2. MGW A is the only source switch and MGW B the only target switch in the re-homing. The forecasted trunks needed for the expansion is shown below:

$$\overline{N}_{Trunk} = \left[Roundup\left(\frac{N_{Trunk}/F_{Redundancy} - B_{IC}}{N_{Card}}\right)\right] \times N_{Card} \quad (6)$$

where $N_{Trunk}$ denotes the forecasted number of trunks in month N. Without a re-home, it's $N_{Trunk-before} = b_n$ from Equation 3. With a re-home, it's $N_{Trunk-after} = B_n$ obtained from Equation 5.

$B_{IC}$ denotes the installed trunks capacity obtained from Table I.

$N_{Card}$ denotes the supportable number of trunks per trunk card in the switch. It varies by vendors' equipments.

$F_{Redundancy}$ is the redundancy factor. It normally ranges 0.7-1.

If $\overline{N}_{Trunk} + B_{IC} \geq B_{Max} \times F_{Redundancy}$, the maximum capacity of the current switch can not suffice the required traffic. A new switch should be added to evenly share the whole required traffic with the current one. The following formula gives the estimated number of new switches needed.

$$n_{New-switch} = \left[Roundup\left(\frac{\overline{N}_{Trunk} + B_{IC}}{B_{Max}}\right)\right] - n \quad (7)$$

where $\overline{N}_{Trunk}$ denotes the total forecasted trunks needed,

International Journal of Next Generation Network (IJNGN), Vol.2, No.1, March 2010
$B_{Max}$ denotes the maximum supportable number of trunks in the switch,
n denotes the number of source or target switches.

As per the re-homing principle 2 in model 1 in section 2, traffic should be always evenly distributed into all the MGWs belong to the same service area. So the required trunks for each new added switch is given by

$$\overline{N}_{New-switch} = \frac{\overline{N}_{Trunk} + B_{IC}}{n_{New-switch} + n} \quad (8)$$

in which $\overline{N}_{New-switch} \leq B_{Max} \times F_{Redunancy}$.

Inputting the known parameters of "before re-homing" and "after re-homing" obtained via Equation 1 into Equation 5 to 8 obtained above, we are able to compare the difference of $\overline{N}_{Trunk}$, $n_{New-switch}$, and $\overline{N}_{New-switch}$ between a "before re-homing" and "after re-homing" scenario, or even between two different re-homing scenarios. If the cost of the trunk, trunk card or switch is further provided from vendors, one can compare the required budget for different scenarios.

## 4. CASE STUDY

The Figure 6 illustrates a re-homing scenario. The re-homing was done in May 2008 since the forecasted trunks (1090) was close to the capacity of current installed trunks (1280). Normally there is a redundancy factor between the actual value and maximum installed value. In our case, the redundancy factor is 85%. So the gap displayed in the figure between the red line and black line shall be always no less than 15% of maximum installed trunks. As per the original plan without a re-homing, new trunks (1470-1280=190) should be added before June 2008 to meet the increase of forecasted trunks. With a re-home, the traffic through the switch is moved out so the number of trunks in the switch was reduced from 1120 to 920 in June 2008. The gap between the black and blue line shows the saved trunks from the re-homing which means it also reduces the loading of the switch and eliminates the need for installing new trunks at this time.

60

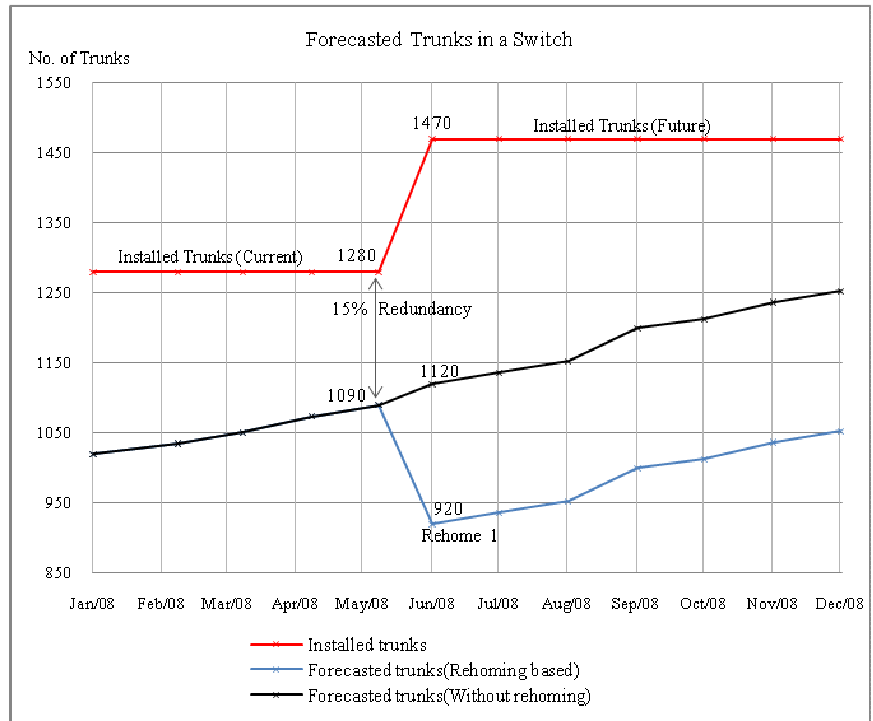

Fig.6. A re-homing analysis

## 5. CONCLUSION

As mobile operators evolve their networks to UMTS or even LTE, they will try to minimize cost and maximize subscriber usage. Therefore, re-homing, as a technique of network optimization, has been widely applied in the radio network by the mobile operators to avoid configuring unnecessary network resources and maintain a high quality of service (QoS) to subscribers. This paper introduced the cell site re-homing in the radio network and analyzed its impact on the core network of GSM/UMTS. All of the re-homing scenarios can be applied to one of the nine proposed re-homing models summarized in the paper. The numerical analysis for the proposed models shows the advantages of the re-homing technique in network optimization and cost savings for the mobile operators.

The current literature provides many applied methods and tools to optimize 2G and 3G radio networks. The core network in this area however has not been thoroughly studied due to its complexity. Network optimization not only helps the mobile operators maintain the existing networks but also plays an important role in evolving the network to 4G. Recently the industry started to talk about the move toward Long Term Evolution (LTE) from 3G. The evolution has been ongoing but it's a lengthy process and requires a systematic and an optimal approach. The high Quality of Service (QoS) of a network ensures a stable and smooth transition from 3G to 4G. Our continuing research will be focused on developing a tool to achieve intelligent and automatic re-homing between switches to replace the current time and labor intensive manual operation for a re-homing.

## REFERENCE

[1] Britvic, V., *Steps in UMTS network design*, Electro-technical Conference, 2004. MELECON 2004.

Proceedings of the 12th IEEE Mediterranean Volume 2, Issue, 12-15 May 2004 Page(s): 461 - 464 Vol.2.

[2] Oom, Jan., Wallentin, Pontus., Rehoming and resource sharing in communications networks, United States Patents:6,738,625.

[3] Shalak, R., Sandrasegaran, K., Agbinya, J., Subenthiran, S, *UMTS core network planning model and comparison of vendor product performance*, Advanced Communication Technology, 2004. The 6[th]International Conference, Volume: 2, page(s): 685- 689.

[4] Harmatos, J, *Planning of UMTS core networks,* Personal, Indoor and Mobile Radio Communications,2002. The 13th IEEE International Symposium, Volume 2, 15-18 Sept. 2002 Page(s):740 - 744 vol.2.

[5] Mishra, A, *Performance characterization of signaling traffic in UMTS core networks*, GlobalTelecommunications Conference, 2003. GLOBECOM '03. IEEE, Publication Date: 1-5 Dec. 2003.

[6] Konstantinopoulou, C.N.; Koutsopoulos, K.A.; Lyberopoulos, G.L.; Theologou, M.E., *Core network planning, optimization and forecasting in GSM/GPRS networks*, Communications and Vehicular Technology, 2000. SCVT-200. Symposium on Volume , Issue , 2000 Page(s):55 – 61

[7] 3GPP TS 25.415, *Technical Specification Group Radio Access Network: UTRAN Iu interface user plane protocols*.

[8] 3GPP TS 25.413, Technical Specification Group Radio Access Network: *UTRAN Iu interface Radio Access Network Application Part (RANAP) signaling*.

[9] 3GPP TS 23.002, Technical Specification Group Services and Systems Aspects; *Network architecture.*

[10] 3GPP TS 25.401, *Technical Specification Group Radio Access Network: UTRAN Overall Description.*

[11] ITU-T I.363.2, *B-ISDN ATM Adaptation Layer Specification: Type 2 AAL Series I: Integrated Services Digital Network - Overall Network Aspects and Functions - Protocol Layer Requirements.*

[12] Vrabel, A.; Vargic, R.; Kotuliak, I., *Subscriber databases and their evolution in mobile networks from GSM to IMS*, ELMAR, 2007, Volume , Issue , 12-14 Sept. 2007 Page(s):115 – 117.

[13] Ouyang, Y. and Fallah, M.H. (2009) 'Evolving core networks from GSM to UMTS R4 version', *Int. J. Mobile Network Design and Innovation*, Vol. 3, No. 2, pp.93–102.

[14] Ouyang, Y. and Fallah, M.H., *A Study of Throughput for Iu-CS and Iu-PS Interface in UMTS Core Network*. Performance, Computing and Communications Conference, 2009. IPCCC 2009. IEEE International. Dec 14-16, 2009.

[15] Ouyang, Y. and Fallah, M.H., *A Study of Throughput for Nb, Mc and Nc Interface in UMTS Core Network*. Performance, Computing and Communications Conference, 2009. IPCCC 2009. IEEE International. Dec 14-16, 2009.

[16] Szekely, I.; Balan, T.; Sandu, F.; Robu, D.; Cserey, S., *Optimization of GSM-UMTS core network for IP convergence in 4G through Mobile IPv6*, Optimization of Electrical and Electronic Equipment, 2008. OPTIM 2008. 11th International Conference on Volume , Issue , 22-24 May 2008 Page(s):217– 222.